\documentclass[12pt]{article}

\usepackage{amsmath,amssymb,amsthm,mathrsfs,bbm}
\usepackage{latexsym,amscd,amsbsy,dsfont,amsfonts}
\usepackage{graphicx}
\usepackage[pdftex,colorlinks=true]{hyperref}

\newcommand{\hs}{\hspace{0.15mm}}

\usepackage{bm}
\usepackage{latexsym}
\usepackage[latin1]{inputenc}
\usepackage{amsfonts}
\usepackage{amssymb}
\usepackage{amsmath}

\numberwithin{equation}{section}

\begin{document}

\title{AdS/CFT beyond the unitarity bound}
\author{Tom\'{a}s Andrade, Donald Marolf   \\
{\it Department of Physics, UCSB, Santa Barbara, CA 93106, USA}}

\maketitle

\begin{abstract}
Scalars in AdS${}_{d+1}$ with squared masses in the Breitenlohner-Freedman window  $-d^2/4 \le m^2 < -d^2/4 +1$ (in units with the AdS scale $\ell$ set to $1$) are known to enjoy a variety of boundary conditions.  For larger masses $m^2 > -d^2/4 +1$, unitarity bounds in possible dual CFTs suggest that such general boundary conditions should lead to ghosts.  We show that this is not always the case as,  for conformally-invariant boundary conditions in Poincar\'e AdS that would naively violate unitarity bounds, the system is generically ghost-free.  Conflicts with unitarity bounds are avoided due to the presence of unexpected pure gauge modes and an associated infrared divergence.  The expected ghosts appear when the IR divergence is removed either by deforming these boundary conditions or considering global AdS.

\end{abstract}

\tableofcontents
\sloppy

\section{Introduction}

The AdS/CFT correspondence relates non-gravitating (boundary) field theories to (bulk) theories which include dynamical gravity and satisfy asymptotically AdS boundary conditions.   For a given bulk field content, such boundary conditions are not unique.  Indeed, deforming any such boundary condition modifies the dual field theory by adding sources for appropriate operators \cite{Maldacena:1997re,Witten:1998qj,Gubser:1998bc,Aharony:1999ti}.

Tachyonic bulk scalars in AdS$_{d+1}$ with squared-masses in the range $m^2_{BF} +1 > m^2 > m^2_{BF}$ near the so-called Breitenlohner-Freedman stability bound $m_{BF}^2 = - d^2/4$   provide particularly interesting examples of this correspondence. Here we have set the AdS length scale $\ell$ to $1$.  As indicated by
\cite{Breitenlohner:1982bm,Breitenlohner:1982jf},  such scalar
fields admit a variety of possible boundary conditions.  In
particular, solutions of the mass $m$ Klein-Gordon equation in AdS are associated with two characteristic fall-off rates near the boundary and
one may set to zero either the faster- or slower-falloff part of
the field. Up to quantum corrections in the bulk ($1/N$ corrections in the CFT, where $N$ is the rank of an appropriate gauge group), the dual CFT operator then has conformal dimension

\begin{equation}
\label{dpm}
\Delta_\pm = \frac{d}{2} \pm \frac{1}{2} \sqrt{d^2 + 4 m^2},
\end{equation}
where the positive (negative) sign corresponds to setting the slow (fast) falloff piece to zero \cite{Balasubramanian:1998sn,Klebanov:1999tb}.  We will refer to such $\Delta_+$ ($\Delta_-$) boundary conditions as Dirichlet (Neumann) below, generalizing terminology that is natural for conformally-coupled scalars (which always reside in the above range). More interesting mixed boundary conditions involving both pieces are also possible \cite{Witten:2001ua,Berkooz:2002ug} and correspond to multi-trace deformations of the Neumann ($\Delta_-$) CFT.  These mixed cases typically break conformal invariance and generate renormalization-group flows between the above two CFT's  -- with an operator of dimension $\Delta_-$ at the UV (ultraviolet) fixed point flowing to an operator of dimension $\Delta_+$ at the IR (infrared) fixed point.  See \cite{Gubser:2002zh,Gubser:2002vv,Hartman:2006dy} for additional evidence in support of this picture.  Similar statements hold at small masses for spin-1/2 fields  \cite{Breitenlohner:1982bm,Breitenlohner:1982jf,Amsel:2008iz} and spin-3/2 Rarita-Schwinger fields \cite{Amsel:2009rr}, as well as for Maxwell fields in AdS${}_4$  \cite{Witten:2003ya,Marolf:2006nd}.

As noted in \cite{Balasubramanian:1998sn,Klebanov:1999tb}, extrapolation of the above picture to larger $m^2$ would lead to $\Delta_- \le (d-2)/2$, violating known unitarity bounds on CFTs
(see e.g. \cite{Minwalla:1997ka}).\footnote{The marginal case $\Delta_- = (d-2)/2$ is allowed only for free fields that do not couple to other operators.}  This suggests that any corresponding bulk theories should contain ghosts.  Indeed, \cite{Compere:2008us} suggested a mechanism through which such ghosts could arise: As observed in \cite{Breitenlohner:1982bm,Breitenlohner:1982jf}, for $m^2 \ge m_{BF}^2 +1$ only the Dirichlet ($\Delta_+$) boundary conditions lead to solutions that are normalizeable with respect to the Klein-Gordon inner product.  More general boundary conditions thus require this inner product to be modified by subtracting suitable counter-terms at the AdS boundary.  Since these counter-terms are designed to cancel a manifestly positive divergence, their sign is negative.  The lack of a manifestly positive definite inner product may then allow the presence of ghosts.   Similar issues with the relevant norms arise for Fermions with large masses \cite{Amsel:2008iz,Amsel:2009rr}, for Maxwell fields in $AdS_{d+1}$ for $d \neq 3$ \cite{Marolf:2006nd}, and for linearized gravitons in any dimension \cite{Compere:2008us}.
In most cases these fields again violate unitarity bounds, though there are interesting exceptions (see section \ref{disc} below).

However, we show below that the expected ghosts do not always arise.  The exception occurs for (conformally-invariant) Neumann ($\Delta_-$) boundary conditions in Poincar\'e AdS, where instead one generically finds  an interesting infrared divergence and unexpected pure gauge modes.   However, the expected ghosts arise when the IR divergence is removed either by deforming these boundary conditions or considering global AdS.

We begin with Neumann boundary conditions in Poincar\'e AdS in section \ref{sec:Neumann} and discuss the above-mentioned IR divergence. We then discuss IR regulators in section \ref{IR regulators} and close with a brief discussion in section \ref{disc}.  We use Lorentz-signature techniques throughout.

\section{Neumann theories in the Poincar\'e patch}
\label{sec:Neumann}

Consider a free scalar field of mass $m^2$ in the Poincar\'e patch of Lorentzian $AdS_{d+1}$, whose line element is given by
\begin{equation}\label{metric PP}
    ds^2 = \ell^2 \left[\frac{dr^2}{r^2} + \frac{1}{r^2} \eta_{ij} dx^i dx^j \right] .
\end{equation}
Here $\eta_{ij}= diag(-+\dots+)$ is the Minkowski metric and $\ell$ is the AdS length, which we henceforth set to $1$. The radial coordinate $r$ is dimensionless with AdS boundary at $r=0$ and the Poincar\'e horizon at $r = \infty$. It is useful to parametrize the mass of the scalar field as
\begin{equation}\label{mass}
    m^2 = m^2_{BF} + \nu^2  ,
\end{equation}
for $\nu \ge 0$. The range in which \cite{Breitenlohner:1982bm,Breitenlohner:1982jf,Balasubramanian:1998sn,Klebanov:1999tb} discuss a variety of boundary conditions is $1 > \nu \ge 0$. In particular, it is at $\nu = \nu_{Unitary} := 1$ that an operator of dimension $\Delta_-$ in a $d$-dimensional CFT saturates the unitarity bound.

Our interest here is in the range $\nu \ge 1$.  As mentioned in the introduction, one obstacle to working in this range is that the Klein-Gordon inner product ceases to be finite unless Dirichlet boundary conditions are imposed.  However, as emphasized in \cite{Compere:2008us}, other `renormalized' inner products may also be considered.  Indeed, the Klein-Gordon current can be associated with the ``bare'' action
\begin{equation}\label{bare action}
    I_0 = - \frac{1}{2} \int_{M} \sqrt{g} [ g^{\mu \nu} \partial_\mu \phi \partial_\nu \phi + m^2 \phi^2] ,
\end{equation}
which diverges for all $\nu \ge 0$ unless Dirichlet boundary conditions are imposed.  As is by now well known (see \cite{Skenderis:2002wp} and references therein), such divergences can be cancelled for all $\nu$ by adding appropriate boundary counter-terms.  For $0 \le \nu < 1$, these counter-terms are algebraic in the fields.  But for $\nu \ge 1$ the required counter-terms contain derivatives along the boundary; e.g., for $1 \le \nu < 2$ one must add an additional counter-term proportional to the integral of $(\partial \phi)^2$.  Since one has then, in effect, added an explicit boundary kinetic term, it is natural to add a corresponding boundary term to the Klein-Gordon inner product.  Indeed, as shown in \cite{Compere:2008us}, such terms are required in order for the total (renormalized) inner product to be conserved.    We refer to \cite{Compere:2008us} for the full prescription relating boundary terms in the action to boundary terms in the inner product.   Below, we proceed at a more intuitive level, checking explicitly that the final renormalized inner product is both finite and conserved.

\subsection{The Action and the norm}

As stated above, we wish to renormalize the bare action (\ref{bare action}) following \cite{Skenderis:2002wp}.  Recall that the equation of motion is just Klein-Gordon equation $\Box_{AdS} \phi = m^2 \phi$, which, in the metric (\ref{metric PP}), reads
\begin{equation}\label{eom in PP}
    r^2 \partial^2_r \phi - (d-1) r \partial_r \phi - m^2 \phi + r^2 \Box_0 \phi = 0 .
\end{equation}
Here $\Box_0$ is the D'Alembertian in the flat boundary metric $\eta_{ij}$.  Near $r=0$, solutions behave as $r^{d/2-\nu}$ and/or $r^{d/2+\nu}$ times some power series in $r^2$.  We will need to keep track of the $[\nu]$ terms in the associated power series which lead to behavior near $r=0$ intermediate between
$r^{d/2-\nu}$ and $r^{d/2+\nu}$.  Here $[\nu]$ denotes the integer parte of $\nu$.  We therefore confine explicit analysis to cases $1< \nu < 2$, though larger (non-integer) values of $\nu$ behave similarly.  The special case $\nu =1$ will be discussed in appendix \ref{marginal}.

For $1 < \nu < 2$, solutions of (\ref{eom in PP}) take the form
\begin{equation}\label{asympt}
    \phi = r^{d/2-\nu} ( \phi^{(0)} + r^2 \phi^{(1)} + r^{2\nu} \phi^{(\nu)} + \ldots  ) ,
\end{equation}
near $r=0$ with the relation
\begin{equation}\label{psi1}
    \phi^{(1)} = \frac{1}{4(\nu-1)}\Box_0 \phi^{(0)} .
\end{equation}
Higher order terms will make no explicit contribution to our calculations below. In terms of (\ref{asympt}), the Dirichlet boundary condition is $\phi^{(0)} = 0$ while the Neumann condition is $\phi^{(\nu)} = 0$.

Using (\ref{asympt}), one may show that the action
\begin{equation}\label{I N}
    I_N = I_0 + \int_{\partial M} \sqrt{\gamma} \left [\rho_\mu \partial^\mu \phi \phi - \frac{1}{2}(d/2 - \nu) \phi^2 + \frac{1}{4(\nu-1)} \gamma^{ij} \partial_i \phi \partial_j \phi \right ] ,
\end{equation}
\noindent where $\rho_\mu$ is the outward-pointing unit normal to $\partial M$, is both finite and stationary on-shell under either Dirichlet or Neumann boundary conditions.  Here $\gamma_{ij} = r^{-2} \eta_{ij}$ is the induced metric on slices of constant $r$ near $r=0$. In particular, a straightforward calculation using (\ref{asympt}), (\ref{psi1}) yields
\begin{equation}\label{d I N}
    \delta I_N = 2 \nu  \int_{\partial M} \phi^{(0)} \delta \phi^{(\nu)},
\end{equation}
under general on-shell variations,
which clearly vanishes when either $\phi^{(0)}=0$ or $\delta  \phi^{(\nu)}=0$.
Finiteness of the on-shell action then follows from finiteness of (\ref{d I N}) since $I_N=0$ for $\phi =0$. The action $I_N$ is also finite off-shell so long as the expansion (\ref{asympt}) and the condition (\ref{psi1}) are imposed as boundary conditions.

We can now read off the renormalized inner product.  We take the bulk Klein-Gordon current associated with a pair of solutions $\phi_1,\phi_2$ to be
\begin{equation}
\label{KGbulk}
j^{bulk}_\mu =  \frac{i}{2} \phi_1^* \stackrel
{\leftrightarrow}
\partial_\mu \phi_2 ,
\end{equation}
and introduce a corresponding boundary current
\begin{equation}
\label{KGbndy}
j^{bndy}_j =  \frac{i}{2} \phi_1^* \stackrel
{\leftrightarrow}
\partial_j \phi_2,
\end{equation}
where   $A \stackrel {\leftrightarrow} \partial B = A \partial B - B \partial A $ and
 the index $j$ ranges only over boundary directions.
The renormalized inner product is then simply
\begin{equation}
\label{ip}
(\phi_1,\phi_2)_{renorm} = (\phi_1,\phi_2)_{bulk} - \frac{1}{2(\nu-1)} (\phi_1,\phi_2)_{bndy},
\end{equation}
where $(\phi_1,\phi_2)_{bulk}, (\phi_1,\phi_2)_{bndy}$ are given by introducing some surface $\Sigma$ with boundary $\partial \Sigma$ at $r=0$, contracting the currents (\ref{KGbulk}), (\ref{KGbndy}) with either the normal $n^\mu $ to $\Sigma$ or the normal $n_\partial^\mu$ to $\partial \Sigma$ within the surface $r=0$, and integrating over $\Sigma$ or $\partial \Sigma$ using the volume measure induced by (\ref{metric PP}).

We will explicitly verify below that (\ref{ip}) is finite as one expects.  However, our main task will be to identify ghosts, associated with violations of unitarity.  For our purposes, it is sufficient to define a ghost to be a mode of definite positive frequency $\omega$
(i.e., $\partial_t \phi = - i \omega \phi$) with negative norm. (The complex conjugate mode has negative frequency and positive norm and may also be called a ghost.)
While the Klein-Gordon norm is positive semi-definite (for positive frequency), one notes that the coefficient of $(\phi_1,\phi_2)_{bndy}$ is negative since $\nu > 1$, as it must be since this term is designed to cancel a positive-signed bulk divergence.   Positivity of (\ref{ip}) (or lack thereof) is thus not manifest and requires
further investigation.

\subsection{Evaluating the norm}
\label{Nnorm}

Although we have focussed on the Neumann boundary condition $\phi^{(\nu)} = 0$, it is useful to compute the norm (\ref{ip}) for general modes satisfying (\ref{asympt}).  This allows us both to consider more interesting boundary conditions in section \ref{IR regulators} and also to check our calculations against standard results for the Dirichlet case.  Our computations below largely follow those of \cite{Compere:2008us} for the analogous graviton modes.

For modes of the form
\begin{equation}\label{basis}
    \phi(r,x) =  e^{ik \cdot x} \psi(k,r) ,
\end{equation}
with boundary momentum $k_j = (\omega, \vec k)$, the radial profile $\psi(k,r)$ satisfies
\begin{equation}\label{eom in PP psi}
    r^2 \psi'' - (d-1) r \psi' - m^2 \psi + r^2 m_{bndy}^2 \psi = 0 ,
\end{equation}
where primes denote radial derivatives and, since we will be most interested in timelike $k^i$, we have defined $m^2_{bndy} = - k^j k_j$ which is positive for timelike modes.  For later purposes, we note that (\ref{eom in PP psi}) can be cast as a Sturm-Liouville (SL) problem with eigenvalue $\lambda = m_{bndy}^2$ for the operator
\begin{equation}\label{general SL}
    L =  r^{d-1}  \left[ - \frac{d}{dr} \left ( r^{1-d} \frac{d}{dr} \right) + m^2 r^{-(d+1)} \right].
\end{equation}

Since $\nu$ is not an integer, a general solution of (\ref{eom in PP psi}) is some linear combination of
\begin{equation}\label{psi D}
    \psi_D (r,k) = 2^\nu m_{bndy}^{-\nu} \Gamma(1+\nu) r^{d/2} J_\nu (m_{bndy} r) = r^{d/2 + \nu} (1 + \ldots),
\end{equation}
and
\begin{equation}\label{psi N}
    \psi_N (r,k) = 2^{-\nu} m_{bndy}^\nu \Gamma(1-\nu) r^{d/2} J_{-\nu} (m_{bndy}r) ,
\end{equation}
where $\psi_D, \psi_N$ satisfy Dirichlet and Neumann boundary conditions respectively as given by (\ref{basis}),(\ref{psi D}), (\ref{psi N}).
We may thus write the general solution of (\ref{eom in PP psi}) for momentum $k$ as
\begin{equation}
\label{psik}
\psi_k  = \phi_k^{(0)} \psi_N(r,k) + \phi_k^{(\nu)} \psi_D(r,k).
\end{equation}
for some constants $\phi_k^{(0)}, \phi_k^{(\nu)}$.
Comparing with (\ref{basis}),(\ref{asympt}) we find $\phi^{(0)} = e^{ik \cdot x} \phi^{(0)}_k$ and $\phi^{(\nu)} = e^{ik \cdot x} \phi^{(\nu)}_k$
for this solution.

Since we include no counter-terms at the horizon or at large $x$, normalizeability at $r = \infty$ and $|x| = \infty$ is determined only by the Klein-Gordon inner product and has the usual implications.  In particular, unless $m_{bndy}$ is real, both $\psi_D, \psi_N$ diverge exponentially at the horizon\footnote{Though under more general (``mixed'') boundary conditions the singular terms from (\ref{psi D}) can cancel against those form (\ref{psi N}).  We will deal with such cases as they arise below.} and neither solution is normalizeable.   We therefore take $m_{bndy}$ real (and define it to be non-negative).
We also take the spatial components $\vec k$ of $k^i = (\omega, \vec k)$ to be  real to avoid divergences as $|\vec x| \rightarrow \infty$.  It then follows that $\omega$ is real as well.

Inserting (\ref{basis}) into (\ref{KGbulk}) we find
\begin{equation}\label{ip bulk}
    ( \phi_1,  \phi_2)_{bulk} = \frac{1}{2} (\omega_1 + \omega_2) (2\pi)^{d-1} \delta(\vec{k}_1 - \vec{k}_2) e^{i(\omega_1 - \omega_2) t} \langle \psi_{1}, \psi_{2} \rangle_{SL} ,
\end{equation}
\noindent where
\begin{equation}\label{rad int D PP}
    \langle \psi_1, \psi_2 \rangle_{SL} = \int_0^\infty dr r^{(1-d)} \psi_{k_1}^*(r) \psi_{k_2}(r) ,
\end{equation}
is the inner product of the Sturm-Liouville problem (\ref{general SL}).  Using (\ref{rad int D PP}) and  (\ref{psi D}), (\ref{psi N}) one sees that as usual the timelike modes are plane-wave normalizeable at the horizon, so for real timelike $k$ we should compute the inner product in an appropriately distributional sense as a function of $m_{bndy}$.  In addition, one finds an interesting result in the lightlike case $m_{bndy} = 0$ for which $\psi_D,\psi_N \propto r^{d/2 \pm \nu}$: while  neither mode is normalizeable at $r= \infty$ for $\nu < 1$, $\psi_N$ becomes normalizeable at the horizon for $\nu > 1$.  Thus, this case cannot be neglected.

For the timelike modes, integrating by parts in the expression $\langle \psi_\lambda, L \psi_\sigma \rangle$ allows one to evaluate (\ref{rad int D PP}) in terms of the Wronskian of the two solutions:
\begin{equation}\label{rad int D PP SL}
    \langle \psi_1, \psi_2 \rangle_{SL} = \frac{r^{(1-d)}}{m_{bndy,1}^2 - m_{bndy,2}^2} [ \psi_1^* \psi_2' -  \psi_2 \psi_1'{}^* ] \big|_0^\infty .
\end{equation}
Here we have used the fact that $m_{bndy,1},m_{bndy,2}$ are continuous parameters and we assume that any singularities at $m_{bndy,1} = m_{bndy,2}$ can be understood in terms of distributions.  We will treat the case where $m_{bndy}$ takes discrete values separately when it arises below.

Let us use the expansion (\ref{asympt}) to examine the terms associated with the boundary $r=0$.
Because the Wronskian is anti-symmetric, the only non-zero contributions come from cross-terms between $\psi_1$ terms and $\psi_2$ terms involving different powers of $r$. Note that the cross term between $\phi^{(0)}$ and $\phi^{(\nu)}$ gives a finite contribution, and that the only other non-vanishing contributions is a divergence associated with the cross term between $\phi^{(0)}$ and $\phi^{(1)}$.  Now, using (\ref{psi1}), we see that the divergent cross term is proportional to the boundary Klein-Gordon norm of $\phi^{(0)}$.  Furthermore, it appears with precisely the right coefficient to be cancelled by the boundary counter-term in (\ref{ip}). As a result,
\begin{equation}\label{ip bulk 2}
    ( \phi_1,  \phi_2)_{renorm} = \frac{1}{2} (\omega_1 + \omega_2) (2\pi)^{d-1} \delta(\vec{k}_1 - \vec{k}_2) e^{i(\omega_1 - \omega_2) t} \langle \psi_{1}, \psi_{2} \rangle_{SL, renorm}
\end{equation}
where
\begin{equation}
\label{SLren}
\langle \psi_{1}, \psi_{2} \rangle_{SL, renorm}=
    \frac{r^{(1-d)}}{m_{bndy,1}^2 - m_{bndy,2}^2} [ \psi_1^* \psi_2' -  \psi_2 \psi_1'{}^* ] \big|^{r = \infty}  + 2 \nu \frac{[(\phi^{(\nu)}_{k_1})^*  \phi^{(0)}_{k_2} -  (\phi^{(0)}_{k_1})^*  \phi^{(\nu)}_{k_2} ]}{m^2_{bndy, 1} - m^2_{bndy, 2}}.
\end{equation}
In particular, the boundary term at $r=0$ vanishes identically for either Dirichlet or Neumann boundary conditions.  Using the fact that the renormalized inner product is conserved (see \cite{Compere:2008us}) and examining the asymptotic expansion near $r=0$ one can show that this feature must persist for higher (non-integer) values of $\nu$. The argument is the same as that given in appendix D of \cite{Compere:2008us}.

It remains to evaluate the first term in (\ref{SLren}).  As in \cite{Compere:2008us}, we do so by introducing a regulator at $r= r_\infty \gg 1$ and taking the limit $r_\infty \rightarrow \infty$ at the end of the calculation.  Recall that for large arguments of $J_\nu$ may be written
\begin{equation}\label{asympt B infinity}
    J_\nu(x) \approx \sqrt{\frac{2}{\pi x}} \cos \left( x - \frac{\pi}{2} \nu - \frac{\pi}{4}  \right)   \,\,\,\,\,\,\, x \gg 1 .
\end{equation}
With the aid of the usual trigonometric identities, one can then write the desired boundary term as a sum of terms involving $\frac{\sin(r_\infty m_{bndy}^\pm)]}{m_{bndy}^\pm}$ and $\frac{\cos(r_\infty m_{bndy}^\pm)]}{m_{bndy}^\pm}$, where  $m_{bndy}^\pm = m_{bndy,1} \pm m_{bndy,2}$, multiplied by coefficients that are independent of $r_\infty$.
One may then take the limit  $r_\infty \rightarrow \infty$ by recalling that, when considered as distributions, we have
\begin{equation}\label{lim sin cos}
\lim_{r_\infty\to\infty} \frac{\sin(r_\infty M)}{\pi M} = \delta(M), \ \ \  \lim_{r_\infty\to\infty} \frac{\cos(r_\infty M)}{\pi M} = 0 .
\end{equation}
Noting that $m_{bndy,1} + m_{bndy,2}$ is positive (and thus that $\delta(m_{bndy,1} + m_{bndy,2})$ vanishes as a distribution) and simplifying the delta-functions by using
\begin{equation}\label{combine deltas}
    \delta(\vec{k}_1 - \vec{k}_2) \delta(m_{bndy,1} - m_{bndy,2}) = \left| \frac{m_{bndy, 1}}{w_1} \right| \delta^{(d)}(k^i_1 - k^i_2) ,
\end{equation}
the $r=\infty$ term in (\ref{SLren}) takes the form:
\begin{equation}\label{ip general bc}
    (\phi_1, \phi_2)^{r=\infty}_{renorm} = (2 \pi)^{d-1} \delta^{(d)}(k^i_1 - k^i_2) \ |\phi^{(0)}_{k_1} C_{\nu,k_1}  + e^{i\pi \nu} \phi^{(\nu)}_{k_1}  C_{-\nu,k_1}|^2,
\end{equation}
where $C_{\nu,k} =2^{-\nu} m^{\nu}_{bndy} \Gamma(1-\nu)$.  Note that (\ref{ip general bc}) is manifestly positive definite.

Finally, we compute inner products with lightlike modes of the Neumann case.  The results above imply that this lightlike mode is orthogonal to any timelike mode.  Since the lightlike radial function is just a monomial ($\propto r^{d/2-\nu}$), it is straightforward to compute the norm using (\ref{ip bulk}) and subtracting the appropriate counter-term as defined by (\ref{ip}). But since $\int_{r_0}^\infty dr r^{1-2\nu} = \frac{r_0^{2(1-\nu)}}{2(\nu-1)}$ the two terms cancel exactly.  The Neumann lightlike mode is a null direction of the norm, and in some sense represents a pure-gauge mode.

In summary, for the Dirichlet and Neumann cases the full inner product is
\begin{eqnarray}\label{ip D final}
    (\phi_1, \phi_2)_D &=& 4^\nu \Gamma^2(1+\nu) (m_{bndy,1})^{-2\nu} (2 \pi)^{d-1} \delta^{(d)}(k^i_1 - k^i_2) |\phi_{k_1}^{(\nu)}|^2, \ \  \ {\rm and}, \\
    \label{ip N final}
    (\phi_1,\phi_2)_N &=& 4^{-\nu} \Gamma^2(1-\nu) (m_{bndy,1})^{2\nu} (2 \pi)^{d-1} \delta^{(d)}(k^i_1 - k^i_2) |\phi_{k_1}^{(0)}|^2,
\end{eqnarray}
which we see differ only by changing the sign of $\nu$. Note that the signs of both (\ref{ip D final}) and (\ref{ip N final}) are manifestly {\it positive} (as is that of the general expression (\ref{ip general bc})). In the Dirichlet case this is just the usual result that the CFT is ghost-free, as expected from the fact that $\Delta_+ > \Delta_{Unitary} := (d-2)/2$.

On the other hand, positivity of the Neumann case may come as a surprise since, given that $\Delta_- < \Delta_{Unitary}$, one naturally expects the theory to contain ghosts.  However, we have already discovered one feature that could allow the theory to evade the usual unitarity bounds: Since the lightlike mode had zero norm, the operator $\phi^{(0)}$  defined by (\ref{asympt}) is not a gauge invariant local operator.  In addition, we will find another (related) novel feature in section \ref{IR div} below.  It is therefore consistent to assume that the above modes are complete though, since the renormalized SL inner product is not manifestly positive definite, we can offer no general proof even of the completeness of the set of radial functions\footnote{One can investigate the existence of solutions with anharmonic dependence on the boundary coordinates such as $t^ne^{ik\cdot x}$. Modes of this form exist only for timelike $k$ and can be constructed from the harmonic modes by taking derivatives with respect to $\omega$.  This means that they are linearly dependent on the above modes and, after smearing against smooth functions of $k$, do not lead to new solutions.}.

It is worth commenting that the key role in the above analysis was played by boundary conditions at the {\it horizon} and at large $|x|$ and not by our detailed calculations.  It was these conditions that forced  $k^i$ to be real and either timelike or null.  From that point, the norm is determined up to an overall constant by Poincar\'e symmetry and scale invariance.  Note that, if one assumed $\phi^{(0)}$ to be gauge invariant, either possible sign would still lead to some tension with the unitarity bounds -- at least at the level of free scalars where one could simply change the sign in front of the action (\ref{I N}) in order to change the sign of the norm.  On the other hand, the positive sign obtained for the Neumann norm did require an explicit calculation.  We note that this sign is critical for stability to be maintained when our scalar is coupled to other fields (such as the bulk graviton).

\subsection{IR divergence}
\label{IR div}

We saw above that the Neumann modes are ghost-free even for $\nu > 1$ where one would expect the bulk to define a dual boundary operator with dimension $\Delta_- < \Delta_{Unitary}$, violating the unitarity bound.  In part, this tension is resolved by the observation that $\phi^{(0)}$ is not gauge invariant. But there turns out to be an additional subtlety which arises when one attempts to compute the two-point function of $\phi^{(0)}$.  In this section, we use a gauge-fixed version of $\phi^{(0)}$ defined by requiring the coefficient of the light-like modes to vanish.  We will see that even this gauge-fixed object does not define an operator of dimension $\Delta_-$.

The key point is an IR divergence that arises when one attempts to construct the two-point function from which one would read off the dimension of the dual operator.  This divergence arises in both the bulk and boundary two-point functions, so that the Neumann theory simply does not exist at the quantum level in the bulk.  We will concentrate on the bulk two-point function, from which one should be able to extract any boundary two-point function by taking an appropriate limit.  However, precisely the same divergence arises if one attempts to construct the boundary two-point function directly.

Assuming that the modes found in section \ref{Nnorm} are complete, the  bulk field operator $\phi$ (gauge-fixed as above) may be expanded as
\begin{equation}\label{mode exp}
    \phi(x,r) = \int_{V^+} d^d k [ a^\dagger (k) u_k(x,r) + a(k) u^*_k(x,r) ] , {\rm with}
\end{equation}
\begin{equation}\label{uk}
    u_k(x,r) = e^{i k_j x^j} \psi_N(r,k) ,\end{equation}
where the integral in (\ref{mode exp}) is over the positive future light-cone  in the tangent space ($k^0 > 0$, or $k_0 < 0$).
The functions $u_k(x,r)$ correspond to positive frequency modes, the frequency being the eigenvalue of the operator $i\partial_t$.    The algebra of the creation/annihlation operators $a$, $a^\dag$ is determined by the norm in the usual way:
\begin{equation}\label{comm}
    [a(k_1), a^\dag(k_2)] = \frac{4^{\nu}}{(2 \pi)^{d-1} \Gamma^2(1-\nu)} (m_{bndy,1})^{-2\nu} \delta^{(d)}(k^i_1 - k^i_2),
\end{equation}
i.e., the right-hand side of (\ref{comm}) would be $+1$ if we had used normalized modes in (\ref{mode exp}).  Defining a vacuum $|0\rangle$ by the condition $a(k) |0\rangle = 0 $ for all $k$ yields the (Wightman) two-point function
\begin{equation}\label{2pt x}
    \langle \phi(x_1,r_1) \phi(x_2,r_2) \rangle = \frac{4^{\nu}}{(2 \pi)^{d-1} \Gamma^2(1-\nu)} \int_{\omega \ge |\vec k|} d \omega d^{d-1} \vec k  e^{ik \cdot(x_1-x_2)} \frac{\psi_N(r_1,k)\psi_N(r_2,k)}{(\omega^2 - |\vec k|^2)^\nu}  .
\end{equation}
For $\nu < 1$, one may check that (\ref{2pt x}) gives the familiar position-space correlator (see e.g. \cite{Klebanov:1999tb}) for an operator of dimension $\Delta_-$. Note, however, that the integrand of (\ref{2pt x}) divergences at $\omega = |\vec k|$ on the light cone in momentum space at all $\omega$.    Thus (\ref{2pt x}) diverges for $\nu > 1$ since for $\omega > 0$ the measure $d\omega d^{d-1} \vec k$ does not degenerate on the light cone. Note that the two-point function of the dual CFT suffers equally from this divergence since it is given by (\ref{2pt x}) with the radial functions $\psi_N (r_,k)$ replaced by constants.  In the pure CFT context, such a divergence was briefly discussed in \cite{Grinstein:2008qk}.

We refer to this divergence as an IR effect since it occurs at $m_{bndy}=0$.  In particular, it is not removed by imposing a cutoff on $\omega$.  However, the fact that it persists for all $\omega$ means that the divergence has UV aspects as well.  For example, one might ask if the divergence can be removed by choosing some other state of the theory (not annihilated by all $a(k)$).  While one can indeed make the two-point function finite for separated points at some fixed time $t$ (by, say, taking the small $m_{bndy}$ modes to be in a squeezed vacuum state) the divergence reappears after any arbitrarily short finite time $\Delta t $.  It appears that the mode-decomposition (\ref{mode exp}) cannot be used to define any non-singular states of the bulk theory\footnote{If one is interested only in globally hyperbolic regions of the bulk, which are thus out of causal contact with the boundary, then the evolution is independent of boundary conditions and one may define quantum states using the Dirichlet modes.}.

In this sense, the theory of a free linear bulk scalar with Neumann boundary conditions simply does not exist for $\nu > 1$. Note, however, that it fails to exist because fluctuations of the field are large.  As a result, any interactions (such as the coupling to gravity) cannot be ignored in a complete analysis.  This raises the interesting question of whether the interacting bulk theory might exist but be intrinsically strongly coupled, though this is beyond the scope of the current paper.  In contrast, section \ref{IR regulators} below explores mechanisms that control the above IR divergence and leave the bulk weakly coupled.

\section{Infrared regulators}
\label{IR regulators}

We now explore two ways of removing the IR divergence of section (\ref{IR div}).   Using intuition from the cases $\nu < 1$ where the theory is well-defined (see \cite{Witten:2001ua}), our first approach is to deform the boundary condition by adding a boundary term to the action $I_N$.  The idea is that this term should contain an operator which, if the dual CFT has existed and if the operator dual to $\phi$ would indeed have had dimension $\Delta_-$, would have been relevant and would have generated a renormalization group flow leading to the Dirichlet theory in the IR.  Our second approach is simply to work in global AdS space where the spectrum of modes is generally discrete and IR divergences of the above form cannot occur.  As we will see, the resulting theories contain ghosts.

\subsection{Deformed boundary conditions}
\label{defBC}

We again focus on the case $1 < \nu < 2.$ The action we choose to explore is
\begin{equation}\label{I mix der}
    I_{\kappa, \lambda} = I_N - \nu \int_{\partial M} d^d x \sqrt{g^{(0)}} [ \kappa \partial_i \phi^{(0)} \partial^i \phi^{(0)} + \lambda (\phi^{(0)})^2  ],
\end{equation}
which is stationary (and finite) with the boundary condition
\begin{equation}\label{bc der}
    \phi^{(\nu)} + \kappa \Box_0 \phi^{(0)} - \lambda \phi^{(0)} = 0 .
\end{equation}
Here $\kappa$ and $\lambda$ are constant of (momentum) dimension $2(\nu-1) > 0$ and $2 \nu > 0$, respectively, whose introduction breaks conformal invariance.

Our main focus below will be to search for ghosts among the modes satisfying the boundary condition (\ref{bc der}). As in section (\ref{Nnorm}), the key question turns out to be whether all modes are timelike; i.e., whether there are modes with finite (renormalized) norm and $m^2_{bndy} \leq 0$. Also as before, this issue is largely determined by normalizeability at the horizon $r = \infty$, which is not affected by the addition of boundary counter-terms (which live at $r=0$). We therefore address this question first, before attempting to compute the details of the renormalized inner product. It will be convenient to analyze separately different regions of the parameter space $(\lambda, \kappa)$.  Below, we again consider modes of definite boundary momentum $k_i$ with $m^2_{bndy} = - k_i k^i$.

{\bf The case $\kappa \neq 0$, $\lambda = 0$:}
For generic complex $m_{bndy}$, the solution grows exponentially at the horizon and is not normalizeable.  Normalizeable solutions can occur only when the growing exponentials cancel between the two terms, in which case $\psi_k$ is proportional to the modified Bessel function of the second kind $K_\nu(pr)$, where to remove factors of $i$ we have introduced $p$ defined by $p^2 = -m^2_{bndy}$ and $\Re \ p >0$.  Thus (for $\nu < 2$) the radial profile for tachyonic solutions is
\begin{equation}\label{euc soln gen nu}
    \psi^E = \frac{2^{1-\nu}}{\Gamma(\nu)} p^\nu r^{d/2} K_\nu(p r) = r^{d/2-\nu} \left [1 + r^2 \frac{p^2}{4(1-\nu)} + r^{2 \nu} 4^{-\nu} \frac{\Gamma(-\nu)}{\Gamma(\nu)} p^{2 \nu} + \ldots \right],
\end{equation}
and, to satisfy (\ref{bc der}) with $\lambda = 0$, we must have
\begin{equation}\label{pT lin mix der gen nu}
    A(\nu) p^{2(\nu-1)} =  \kappa
\end{equation}
\noindent where $A(\nu) := 4^{-\nu} \frac{\Gamma(-\nu)}{\Gamma(\nu)} $. Since $\Re p > 0$, let $p = Re^{i\theta}$ for $R> 0$, $|\theta| < \pi/2$ and note that for $1 < \nu < 2$ we have $|2(\nu-1) \theta| < \pi$.  Moreover, because $\Gamma(-\nu) > 0$ in this regime, $A(\nu)$ is positive.  Thus (\ref{pT lin mix der gen nu}) has no solutions for negative $\kappa$ and one solution for positive $\kappa$. The $\kappa > 0$ solution has real $p$, or $m^2_{bndy}< 0$, and is a tachyon.

The analysis of the timelike and lightlike modes at the horizon proceeds as in the Neumann case. In particular, we note that since the lightlike mode has $\Box_0\phi^{(0)}=0$, this mode {\it is} the Neumann lightlike mode for all $\kappa$.  However,
because our $\kappa$-deformation is again a boundary kinetic term, the conserved norm differs from the inner product (\ref{ip})
by an an explicit finite boundary term proportional to the Klein-Gordon norm of $\phi^{(0)}$, which, in view of (\ref{I mix der}), comes with the coefficient $2 \nu \kappa$. It follows that the lightlike modes are normalizeable and have positive norm for $\kappa > 0$, though they are ghosts for $\kappa < 0$.

Computing inner products of the timelike modes is also straightforward.  Since $m_{bndy}$ is continuous, we may again evaluate (\ref{ip}) using (\ref{SLren}). In doing so, due to the boundary condition (\ref{bc der}), the $r = 0$ term in (\ref{SLren}) also gives a finite contribution.  As one might expect, these contributions cancel exactly and the SL product is given entirely by the positive-definite contribution (\ref{ip general bc}) at the horizon.

For $\kappa > 0$ we must also compute the norm of the tachyonic mode.  For simplicity, consider modes with real $\omega > 0$ such as occur for large enough spatial momenta $\vec k$. The properties of the general case are related by Lorentz invariance.
Due to the exponential decay of the tachyon at $r\rightarrow \infty$, we see from (\ref{SLren}) that it is orthogonal to the timelike modes  (since, as above, the terms in (\ref{SLren}) at the AdS boundary cancel against the explicit new boundary term in the norm).  The norm of the tachyon can also be computed using (\ref{SLren}) by considering two profiles of the form (\ref{euc soln gen nu}) for general $p_1,p_2 > 0$ and then taking the limit where $p_1,p_2 \rightarrow p$ as defined by (\ref{pT lin mix der gen nu}).  The exponential decay at the horizon again means that the contribution from $r=\infty$ vanishes.  However, since the boundary condition (\ref{bc der}) does not hold for general $p_1,p_2$, the $r=0$ term in (\ref{SLren}) no longer entirely cancels against the explicit new boundary term.  Instead, we find that the Sturm-Liouville part of the renormalized inner product is
\begin{equation}
\langle \psi^E,\psi^E \rangle =  2 \nu \kappa (1- \nu),
\end{equation}
so that the tachyon is a ghost when it exists (i.e., for $\kappa > 0$). Therefore, we conclude that the theory defined by (\ref{I mix der}) and (\ref{bc der}) contains ghosts all values of $\kappa \neq 0$ with $\lambda = 0$.

{\bf The case $\kappa = 0$, $\lambda \neq 0$:}
For non-zero $\lambda$, the lightlike modes are not normalizeable at the horizon.  This can be seen from the fact that (\ref{bc der}) requires $\phi^{(0)} \neq 0$, so the radial profile behaves like $r^{d/2+\nu}$ at the horizon.

However, we find non-timelike modes for all $\lambda$. To see this, note that the boundary condition (\ref{bc der}) is now
\begin{equation}\label{bc der2}
    A(\nu) p^{2 \nu}  =  \lambda ,
\end{equation}
with $A(\nu)$ defined as above (recall that $A(\nu) > 0$ for $1 < \nu < 2$).
For any $\lambda > 0$ (\ref{bc der2}) has a solution with real $p$. Using the technique explained above to compute inner products for modes lying in the discrete part of the spectrum, one finds the corresponding  SL norm to be
\begin{equation}\label{SL norm re l>0}
    \langle \psi, \psi \rangle = - 2 \frac{\nu^2 \lambda}{p^2} ,
\end{equation}
so this tachyon is a ghost.

For $\lambda < 0$, (\ref{bc der2}) has a pair of complex conjugate solutions $p,p^*$. Denoting the corresponding modes by $\psi_1, \psi_2$, the matrix of their SL products can be written
\begin{equation}\label{SL matrix}
    \langle \psi_i, \psi_j \rangle = \left(
                                       \begin{array}{cc}
                                         0 & a \\
                                         a^* & 0 \\
                                       \end{array}
                                     \right)
\end{equation}
\noindent where $a$ is a non-vanishing complex number.  Since the eigenvalues of (\ref{SL matrix}) are given by $\pm |a|$, after making appropriate modifications to (\ref{ip bulk 2})  to allow for complex frequencies, one finds solutions with negative norm and $\Re \omega > 0$ (which we refer to as complex ghosts).

{\bf The case $\kappa \neq 0$, $\lambda \neq 0$:}
The situation is not alleviated by turning on a non-zero $\kappa$. The boundary condition is
\begin{equation}\label{bc der lambda}
   A(\nu) p^{2 \nu}  =  \kappa p^2 + \lambda .
\end{equation}
For $\lambda > 0$, there is a solution with real $p$ for any $\kappa$.  The corresponding SL norms is
\begin{equation}\label{SL norm re l>0}
    \langle \psi, \psi \rangle = - 2 \nu \left[ \kappa(\nu - 1) + \frac{\nu \lambda}{p^2} \right] = -2\nu^2 A(\nu)p^{2(\nu-1)} + 2 \nu \kappa.
\end{equation}
The first expression in (\ref{SL norm re l>0}) is manifestly negative for $\kappa > 0$ while the second is manifestly negative for $\kappa < 0$.

Consider now $\kappa > 0$, $\lambda < 0$ and introduce
\begin{equation}\label{redef 1}
    \kappa = A(\nu) \kappa_0^{2 \nu - 2} \,\,\,\,\,\,\, \lambda = - A(\nu) \lambda_0^{2 \nu} \,\,\,\,\,\,\, \alpha^2 = \left( \frac{\lambda_0}{\kappa_0} \right)^{2 \nu} \,\,\,\,\,\, \hat{p} = \frac{p}{\kappa_0} .
\end{equation}
We may then write (\ref{bc der lambda}) as
\begin{equation}\label{bc der hat}
   \hat{p}^{2 \nu}  =  \hat{p}^2 - \alpha^2 .
\end{equation}
For $\alpha^2 < \alpha^2_0 = \nu^{1/(1-\nu)} - \nu^{\nu/(1-\nu)}$ there are two real solutions $p_1,p_2$ satisfying $p_1^2 \le \frac{\nu \alpha^2}{\nu-1}$ and $p_2^2 \ge \frac{\nu \alpha^2}{\nu-1}$. The SL norm of these tachyons is given by (\ref{SL norm re l>0}), which in terms of the dimensionless variables of (\ref{redef 1}) reads
\begin{equation}\label{sl norm 2}
    \langle \psi, \psi \rangle = 2 \nu \kappa \left[ (1-\nu) + \nu \frac{\alpha^2}{\hat{p}^2}  \right].
\end{equation}
Thus $p_2$ corresponds to a ghost (while $p_1$ does not). For $\alpha^2 > \alpha_0^2$ the solutions $p_1,p_2$ become complex.  As a result, a linear combination is again a ghost by the same argument as above.  For the marginal case $\alpha^2 = \alpha^2_0$, the tachyons coincide and the norm vanishes. However, by taking appropriate limits as $\alpha \rightarrow \alpha_0$ one can construct a new (linearly independent) mode with anharmonic time dependence.  While we have not analyzed this case in detail, both continuity and analogy with known cases (such as the famous logarithmic mode \cite{Grumiller:2008qz}, \cite{Andrade:2009ae}, associated with chiral gravity \cite{Li:2008dq}, \cite{Maloney:2009ck}) suggests that some linear combination is again a ghost.

It remains to study the region $\lambda < 0$, $\kappa < 0$. While there are no real solutions, a pair of complex conjugate solutions can always be found.  As above, these are complex ghosts.

\subsection{Global AdS}
\label{global}

Another way to regulate the IR divergence of section \ref{IR div} is to consider global AdS space, where the spectrum is discrete. We take the metric to be
\begin{equation}\label{ds2 AdS global}
    ds^2 = \sec^2 \rho ( -dt^2 + d \rho^2 ) + \tan^2 \rho d \Omega_{d-1}
\end{equation}
so that the origin is $\rho = 0$ while the boundary is $\rho = \pi/2$.  Here $d \Omega_{d-1}$ is the line element in $S^{d-1}$. Below, we consider the theory with Neumann boundary conditions $\phi^{(\nu)}=0$ in this setting and find that theory contains ghosts. For comparison, we also include the standard discussion of Dirichlet boundary conditions \cite{Breitenlohner:1982bm}, \cite{Breitenlohner:1982jf}, \cite{Balasubramanian:1998sn}. As usual, we parametrize the mass as $m^2 = -d^2/4 + \nu^2$ and consider $1 < \nu < 2$.

In these coordinates, the Klein-Gordon equation reads
\begin{equation}\label{KG global rho}
   [ - \cos^2 \rho \partial^2_t + \cos^2 \rho \partial^2_\rho + (d-1) \cot \rho \partial_\rho + \cot^2 \rho \nabla^2_\gamma ] \phi = m^2 \phi.
\end{equation}
We decompose the scalar field as $\phi = e^{-i \omega t} Y_{\vec{\ell}}(\Omega) F(\rho)$, where $Y_{\vec{\ell}}(\Omega)$ are spherical harmonics satisfying $\nabla^2_\gamma Y_{\vec{\ell}} = - \ell (\ell + d-2) Y_{\vec{\ell}}$, where the principal angular quantum number, $\ell$, is a non-negative integer.

We now proceed to determine the Dirichlet and Neumann spectrum closely following \cite{Balasubramanian:1998sn}. Let us first find the solutions that are well behaved at the origin. To do so, it is convenient to make the change $y= \sin^2 \rho$. In terms of this variable, the two linearly independent solutions of (\ref{KG global rho}) can be written in terms of Hypergeometric functions as
\begin{equation}\label{phi1 global}
    \phi^{(1)}(y) = (1-y)^{\frac{d}{4} - \frac{\nu}{2}} y^{\ell/2} \hs _2F_1 \left \{ \frac{1}{4} [ d+ 2(\ell+\omega-\nu) ] , \frac{1}{4} [ d+ 2(\ell-\omega-\nu) ] , \ell + \frac{d}{2}, y \right \}
\end{equation}
\begin{eqnarray}\label{phi2 global}
    \phi^{(2)}(y) &=& (1-y)^{\frac{d}{4} - \frac{\nu}{2}} y^{1- \frac{d}{2} - \frac{\ell}{2}} \\
\nonumber
     & & _2F_1 \left \{ -\frac{1}{4} [d+2(\ell-\omega+\nu-2)] , -\frac{1}{4} [d+2(\ell+\omega+\nu-2)] ,\frac{1}{2} (4 - d - 2\ell) , y \right \}
\end{eqnarray}
The functions (\ref{phi1 global}) and (\ref{phi2 global}) are linearly independent if $\ell + d/2$ is not an integer, i.e. if $d$ is odd. If $d$ is even, we can use a basis of solutions consisting on $\phi^{(1)}$ and another solution with a logarithmic branch.   For odd $d$ we may expand (\ref{phi1 global}) and (\ref{phi2 global}) near the origin to find that only $\phi^{(1)}$ is regular.  For even $d$, the logarithmic solution is also divergent at the origin and so again $\phi^{(1)}(y)$ is the relevant solution.

We now study the behavior of (\ref{phi1 global}) near the AdS boundary. To do so, it is useful to first solve the wave equation using the radial variable $z = \cos^2 \rho$, so the boundary is at $z=0$. The solutions can be written as
\begin{equation}\label{phi plus}
    \phi^{(+)}(z) =  (1-z)^{\ell/2} z^{\frac{d}{4} + \frac{\nu}{2}} \hs _2 F_1 \left \{ \frac{1}{4} [ d + 2(\ell-\omega+\nu)], d + 2(\ell-\omega+\nu), 1+\nu, z  \right\},
\end{equation}
\begin{equation}\label{phi minus}
    \phi^{(-)}(z) =  (1-z)^{\ell/2} z^{\frac{d}{4} - \frac{\nu}{2}} _2 F_1 \left \{ \frac{1}{4} [ d + 2(\ell-\omega-\nu)], d + 2(\ell-\omega-\nu), 1-\nu, z \right\}.
\end{equation}
Since $\nu$ is not an integer the solutions (\ref{phi plus}), (\ref{phi minus}) are linearly independent.  At small $z$ we have
\begin{equation}\label{phi plus asympt}
    \phi^{(+)}(z) \sim z^{\frac{d}{4} + \frac{\nu}{2}} (1 + O(z)) ,
\end{equation}
\begin{equation}\label{phi minus asympt}
    \phi^{(-)}(z) \sim z^{\frac{d}{4} - \frac{\nu}{2}} (1 + O(z)) .
\end{equation}
The subleading terms in the series expansions (\ref{phi plus asympt}) and (\ref{phi minus asympt}) are integer powers of $z$,
so $\phi^{(+)}$ satisfies Dirichlet boundary conditions while $\phi^{(-)}$ satisfies the Neumann condition.

As in \cite{Balasubramanian:1998sn}, we write $\phi^{(1)}$ in terms of $\phi^{(\pm)}$ using a standard identity (see, e.g. \cite{Abramowitz1965}) that relates a Hypergeometric function of argument $y=\sin^2 \rho$ to a pair of Hypergeometric functions of argument $1-y= \cos^2 \rho$. We find
\begin{equation}\label{phi1 phi mp}
    \phi^{(1)}(\sin^2\rho) = C_+ \phi^{(+)} (\cos^2\rho) + C_- \phi^{(-)} (\cos^2\rho) ,
\end{equation}
\noindent where
\begin{equation}\label{cplus}
   C_+ = \frac{\Gamma(\ell + \frac{d}{2}) \Gamma(-\nu)}{\Gamma( \frac{1}{4}( d- 2 \nu + 2(\ell+\omega)) ) \Gamma( \frac{1}{4}( d- 2 \nu + 2(\ell-\omega)) ) } ,
\end{equation}
\begin{equation}\label{cminus}
    C_- = \frac{\Gamma(\ell + \frac{d}{2}) \Gamma(\nu)}{\Gamma( \frac{1}{4}( d+ 2 \nu + 2(\ell-\omega)) ) \Gamma( \frac{1}{4}( d+ 2 \nu + 2(\ell+\omega)) ) } .
\end{equation}
As a result, while Dirichlet boundary conditions would require $C_- = 0$, our Neumann conditions require $C_+ =0$.  I.e., the Dirichlet ($\omega_D$) and Neumann ($\omega_N$) frequencies satisfy
\begin{eqnarray}\label{wD}
    \omega_D &=& \pm \left [ \ell + 2 n + \left( \frac{d}{2} + \nu \right )   \right] \,\,\,\,\,\,\,\,\,\,\, n = 0, 1, 2, \ldots, \\
    \omega_N &=& \pm \left [ \ell + 2 n + \left( \frac{d}{2} - \nu \right )   \right] \,\,\,\,\,\,\,\,\,\,\, n = 0, 1, 2, \ldots.
\end{eqnarray}
We emphasize that this is the complete mode spectrum. Although a priori we could have found frequencies anywhere in the complex plane, the analytic structure of $\Gamma$-functions implies that all solutions have real $\omega$.  In particular, the theory has no instabilities for either Dirichlet or Neumann boundary conditions.

The renormalized inner product may be calculated using the method described in section (\ref{defBC}) for modes in the discrete spectrum of the Sturm-Liouville operator.  The result for Dirichlet modes (normalized so that $\phi^{(\nu)}_k =1$) is
\begin{equation}\label{ip global D norm}
   (\hat{\phi}_I, \hat{\phi}_J)_{D} = \delta_{IJ} \pi (-1)^{n+1} n! \frac{\csc(\pi \nu) \Gamma(d/2+\ell+n) \Gamma(-n-\nu) }{\Gamma(d/2+\ell+n+\nu)\Gamma^2(-\nu)} .
\end{equation}
where $I, J$ denote collectively the frequency and angular momentum of the modes, while for  Neumann modes (normalized so that $\phi^{(0)}_k =1$) we obtain
\begin{equation}\label{ip global N norm}
   (\hat{\phi}_I, \hat{\phi}_J)_{N} = \delta_{IJ} \pi (-1)^{n+1} n! \frac{\csc(-\pi \nu) \Gamma(d/2+\ell+n) \Gamma(-n+\nu) }{\Gamma(d/2+\ell+n-\nu)\Gamma^2(\nu)},
\end{equation}
which is just (\ref{ip global D norm}) with $\nu$ replaced by $-\nu$.   One may check that (\ref{ip global N norm}) agrees with (\ref{ip N final}) by taking the limit of large $m_{bndy}^2$ (defined by the eigenvalue of $\Box_0$) and using Stirling's approximation.

As expected, the Dirichlet norm (\ref{ip global D norm}) is positive definite. However, (\ref{ip global N norm}) can become negative for small values of $n$ (recall that we consider $1 < \nu < 2$).   For $\ell \ge 1$, the $n=0$ mode is a ghost for any $\ell$.   For $d > 2\nu$, this is also true of the mode with $n=0, \ell =0$.  For $d < 2\nu$ (which is always true for $d=2$ and can also arise in our range of $\nu$ for $d=3$) the $n > 0, \ell =0$ modes are ghosts (though the $n=0, \ell=0$ mode is not).  The case $d = 2\nu$ (i.e., $d=3, \nu = 3/2$) requires special treatment, but again contains ghosts, e.g. for $n=0$, $\ell \geq 1$.

It is interesting to note that (while there are various exceptions) these ghosts are generally tachyons.  In particular, for Neumann boundary conditions we have
 \begin{equation}\label{M2 bndy N}
    m^2_{bndy} = (d/2 + 2n - \nu)^2 + 2\ell(1+ 2n -\nu).
\end{equation}
Setting $n=0$ one finds $m_{bndy}^2 < 0$ for $\ell > \frac{(d/2-\nu)^2}{2(\nu-1)}$.

\section{Discussion}
\label{disc}

Our main result was to describe the treatment of non-Dirichlet boundary conditions for scalar fields in AdS${}_{d+1}$ with squared masses $m^2 > m^2_{BF} +1$; i.e., above the Brietenlohner-Freedman window for which this had previously been explored.  In this range, in analogy with the treatment of gravitons in \cite{Compere:2008us}, the norm required counter-terms associated with boundary counter-terms in the action that contain derivatives.  Naive extrapolation of the results for smaller masses suggests that these theories should contain ghosts.  This was almost true, but conformally invariant (Neumann) boundary conditions
for non-integer $\nu = \sqrt{m^2 -  m^2_{BF}}$ are an interesting technical exception.  In these special cases, correlators of the bulk quantum field $\phi$ are IR divergent even at finite separation.   The same IR divergence appears in correlators of candidate dual CFT operators.   In addition, the theory contains pure gauge modes so that the Neumann boundary operator $\phi_N$ that one expects to violate the unitarity bound is not in fact gauge-invariant.

Since correlators of local gauge invariants (such as $\Box_0 \phi$) turn out to be well-defined despite the IR divergence, one might take the perspective that the {\it theory} is well-defined but that our gauge-fixed $\phi$ is not a good operator.  However, noting that the stress tensor is not gauge-invariant in the above sense, it seems difficult to maintain this perspective in the presence of interactions.  In contrast, since the IR divergence makes fluctuations in $\phi$ large, it remains possible that including appropriate interactions could tame this divergence and render correlators of $\phi$ finite.  Note that the resulting bulk AdS theory would be intrinsically strongly-coupled.

Instead of following such a path, we investigated IR modifications which make correlators of $\phi$ well-defined while leaving the theory weakly coupled in the bulk.  Such modifications remove the pure gauge modes but also introduce tachyonic ghosts\footnote{Had the tachyons not appeared, our IR modifications would have defined some renormalization-group flow that approached the Dirichlet theory in the IR.  But presumably it would not have approached the Neumann theory in the UV.  We see no reason why some ghost-free theory of this type should not exist.  While it remains a challenge to give a general prescription in the AdS/CFT context, we note that section 3.3 of \cite{Faulkner:2010gj} gives an example where the Dirichlet theory lives in AdS${}_2$ with positive $m^2$.}.  Our treatment was explicit for the case $1 \le \nu < 2$ (with the special case $\nu =1$ treated in appendix \ref{marginal}), but the behavior at larger $\nu$ is similar.

The positive definite norm for Neumann boundary conditions and non-integer $\nu$ is directly analogous to that found in \cite{Compere:2008us} for linearized gravitons when $d$ is odd.  What we now see is that such gravitons are also associated with extra `pure gauge' states with lightlike momentum and an IR divergence in the gauge-fixed two-point function.  Note that these `extra' pure gauge modes mean that even objects like the linearized Weyl tensor (which is usually an observable in linearized gravity) is not in fact gauge invariant.  We also see that ghosts arise when conformal invariance is broken.  Indeed, it is clear from section \ref{IR regulators} that a bulk analysis of the relevant ghosts will yield results for odd $d$ analogous to those reported in \cite{Hawking:2000bb} for $d=4$ using different techniques.

One can again associate the above issues for gravitons with violations of unitarity bounds (see e.g. \cite{Minwalla:1997ka}).  One subtlety is that, as shown in \cite{Compere:2008us}, for Neumann boundary conditions the boundary metric is not a local gauge-invariant operator as diffeomorphmism are now pure gauge.  But at least at the level of linearized gravity (where our analysis is actually performed), the linearized Ricci $R_{ij}$ and Weyl tensors $C_{ijkl}$ of the boundary metric are gauge invariant (or would be, if the extra pure gauge modes with lightlike momentum did not arise).  In particular, for $d > 2$ the traceless part of $R_{ij}$ is a spin-2 operator of dimension 2, where we have ignored certain logarithms that appear for odd $d$ and break conformal invariance.  Thus the fact that the unitarity bound for spin-2 operators is $d$ leads one to expect ghosts.

There are, however,  further subtleties for small $d$. For $d=2$ we see no tension with unitarity bounds, though the Neumann theory is a Liouville CFT with the usual negative central charge \cite{Compere:2008us}.  For $d=3$, the fact that the pure Neumann theory does not contain the expected ghosts is required by electromagnetic-duality for linearized gravitons, which is related to the change from Dirichlet to Neumann boundary conditions; see \cite{Bakas:2008gz,deHaro:2008gp,Mansi:2008bs,Leigh:2003gk,Leigh:2003ez,Bakas:2008zg}.  In this context, the appearance of new gauge modes does signal that only operators insensitive to these modes are well-defined.  In particular, the boundary stress tensor of the dual magnetic theory is related to third derivatives of the (electric) boundary metric operator and has finite two-point functions. However, as for the scalar case, it is difficult to see how this picture can be extended beyond the linearized level.

For completeness, let us also discuss the situation for Maxwell fields (whose details are given in appendix \ref{highS}).  Under Neumann boundary conditions for $d \ge 3$, the operator dual to the bulk potential $A_\mu$ is a boundary gauge field $A_i$, but the associated field strength  $F_{ij}$ would be gauge invariant in the absence of our extra pure gauge modes.  Scalings of the mode functions suggests that this $F_{ij}$ has dimension $2$ for all $d$.  For $d > 4$, $F_{ij}$ violates the associated unitary bound $\max(d-2,2) = d-2$ \cite{Minwalla:1997ka}\footnote{See \cite{ElShowk:2011gz} for the explicit expression of the bound in terms of $d$ and a recent discussion of gauge fields in scale invariant theories.}, where we have again ignored the fact that logarithms break conformal invariance for even $d$.

But again there are subtleties for small $d$.  For $d=4$, it turns out that $F_{ij}$ saturates the unitarity bound.  This means that unitarity would also imply $\partial^i F_{ij} =0$; i.e., it  would require Maxwell's equations to be satisfied on the boundary.  But this condition is clearly not fulfilled due to the existence of timelike modes.  For $d=3$, $F_{ij}$ is dual to a current $j^k = \epsilon^{kij} F_{ij}$.  This again saturates a unitarity bound and requires that $j^k$ be conserved.  But now $\partial_k j^k$ vanishes due to the Bianchi identity , so there is no need for ghosts and indeed none arise (see \cite{Ishibashi:2003jd,Witten:2003ya,Marolf:2006nd}).  The interesting case is $d=2$, where the relevant operator in the Neumann theory is a conserved current\footnote{For $d=2$ Maxwell fields, the more familiar Dirichlet boundary conditions lead only to a boundary gauge field $A_i$ \cite{Marolf:2006nd}.} and so satisfies all unitarity bounds.  Despite this fact, the theory contains a ghost.  The detailed analysis is presented in appendix \ref{highS}, but is also clear from the fact that the bulk AdS${}_3$ theory is dual to that of a massless scalar which has $\nu =1$ and so contains tachyonic ghosts.

To provide a more familiar perspective, it is useful to rephrase this discussion purely in terms of a supposed dual field theory.  For simplicity, let us focus on the $d=4$ case of Maxwell theory in AdS${}_5$.  Recall \cite{Compere:2008us} that passing to Neumann or mixed boundary conditions is equivalent to coupling the dual CFT to a dynamical Maxwell field.  The presence of logarithms means that this theory is not conformal.  Indeed, after including appropriate counter-terms a scale transformation shifts the coefficient of the Maxwell kinetic term on the boundary (the analogue of the term $(\partial \phi^{(0)})^2$ in \ref{I mix der}).  I.e, the coefficient of this kinetic term runs logarithmically and always takes the `wrong' sign at sufficiently small scales.  Rescaling the boundary Maxwell field in the usual way, one sees that this is just the expected running of the coupling constant, which diverges at some scale and then becomes imaginary.  In other words, in this context our effect is just a large $N$ version of the Landau pole of QED (see e.g. \cite{Weinberg:1996kr}) and the ghosts arise from attempting to extend the theory further into the UV beyond the pole.  We presume that a similar language could be used to describe the other cases as well, perhaps using the technology of \cite{Heemskerk:2010hk,Faulkner:2010jy}.

\section*{Acknowledgements}
We thank Ofer Aharony, David Berenstein and Joe Polchinski for conversations concerning unitarity bounds in CFTs. We also thank Mukund Rangamani for encouraging this line of investigation and Tom Faulkner, Juan I. Jottar, Rob Leigh, Mauricio Romo, Jorge Santos and Mark Srednicki for other discussions. Our work was supported in part by the US National Science Foundation under grant PHY08-55415 and by funds from the University of California. In addition, TA was partly supported by a Fulbright-CONICYT fellowship. T.A. is also pleased to thank University of Illinois at Urbana-Champaign for their hospitality during the completion of this work.

\appendix

\section{The special case $\nu=1$}
\label{marginal}

This appendix studies the special case $\nu=1$, or $m^2 = m^2_{BF}+1$. Much of the calculations follow in the same way as in the generic case, so we only summarize the main points below.  Other special cases of integer $\nu > 1$ should behave similarly.

 Consider the action
\begin{equation}\label{I nu=1 mix der}
    I = I_0 + \frac{1}{2}\int_{\partial M} \sqrt{\gamma} \left[ (d/2-1) \phi^2  + \frac{1}{2} (\log r^2 + c) \gamma^{ij} \partial_i \phi \partial_j \phi \right]
\end{equation}
Using the asymptotic expansion
\begin{equation}\label{asympt m1}
    \phi = r^{d/2-1}( \phi^{(0)} + r^2 \phi^{(2)} + r^2 \log r^2 \phi_{log}  ), \ \ \ {\rm with} \ \ \
    \phi_{log} = - \frac{1}{4} \Box \phi^{(0)},
\end{equation}
one may verify that (\ref{I nu=1 mix der}) is finite and satisfies
\begin{equation*}
    \delta I = -2 \int_{\partial M} [ \phi^{(2)} + (1 - c)\phi_{log} ] \delta \phi^{(0)}.
\end{equation*}
This action is thus appropriate for either Dirichlet boundary conditions ($\phi^{(0)}=0$) or for $\phi^{(2)} + (1 - c)\phi_{log} =0$. Note that any finite value of $c$ is related to $c=0$ by an AdS isometry that rescales $r$.  It therefore suffices to analyze the case $c=0$ on which we focus below.

Let us begin by examining the case with $m_{bndy}$ not real and defining $p^2 = - m_{bndy}^2$ with $\Re p > 0$.  Setting $\phi^{(0)}_k =1$ as usual, regularity at the horizon requires
\begin{equation}\label{Euc soln nu =1 asympt}
    \psi = p r^{d/2} K_1(pr) = r^{d/2-1}\left[ 1 + \frac{p^2}{4} \log r^2 + r^2 \frac{p^2}{4}\left( -1 + 2 \gamma + \log(p^2/4)  \right) + \ldots \right]
\end{equation}
\noindent where $\gamma$ is the Euler-Mascheroni constant. The ($c=0$) boundary condition is satisfied only for $p = 2e^{-\gamma}$. The radial profile of the lightlike modes ($m_{bndy}=0$) is simply given by $\psi = r^{d/2-1}$, so they are not normalizeable at the horizon due to a logarithmic divergence. In contrast, the timelike modes (with real $m_{bndy} > 0$) are all plane-wave normalizeable.

The inner products again take the form (\ref{ip bulk}) plus a counter-term, and thus reduce to computing a renormalized Sturm-Liouville inner product:
\begin{equation}\label{ip N tach nu = 1}
    \langle \psi_1, \psi_2 \rangle_{SL, renorm} = \int_0^\infty r^{1-d} \phi_1 \phi_2 dr  + \frac{1}{2}\log r^2 r^{2-d} \phi_1 \phi_2 \big |_{r=0}.
\end{equation}
The calculation proceeds are before for the timelike modes and the tachyon. Fixing $\phi^{(0)}_k =1$ for all modes, we find
\begin{eqnarray}\label{ip N nu = 1}
    (\phi_1, \phi_2) = \frac{m_{bndy}^2}{4}[(2\gamma + \log(m_{bndy}^2/4))^2 + \pi^2 ] (2 \pi)^{d-1} \delta^{(d)}(k_1 - k_2) , \ \ \ k \  {\rm timelike} \cr
    (\phi_1, \phi_2) = - \frac{\omega}{2} (2 \pi)^{d-1} \delta^{(d-1)}(\vec k_1 - \vec k_2) , \ \ \ {\rm for \ tachyons \  with \ real} \ \omega.
\end{eqnarray}
Once again the timelike modes are ghost-free but the tachyon is a ghost. From the analysis above it follows trivially that double trace deformations of the form $\int_{\partial M} (\partial \phi^{(0)})^2$ cannot cure this pathology since they simply correspond to considering $c \neq 0 $ which is physically equivalent to $c=0$. Furthermore, we have verified explicitly that turning on a boundary mass term in addition to the aforementioned deformation also leads to ghosts.

\section{Maxwell fields}
\label{highS}

We consider a free Maxwell field in a fixed (Poincar\'e patch) $AdS_{d+1}$ background with action,
\begin{equation}\label{I=I0+B}
    I = -\int_{M} \frac{1}{4} \sqrt{g} F^{\mu \nu}F_{\mu \nu} + B,
\end{equation}
\noindent where $B$ is a boundary term that is chosen such that the action is finite and that  has an extremum when the boundary conditions are imposed.  We wish to impose Neumann-like boundary conditions as will be discussed below.  The equations of motion are of course
    $\nabla_\mu F^{\mu \nu} = 0$. Choosing the radial gauge $A_r = 0$, the $r$ component of reads
\begin{equation}\label{Mr}
    \partial_r \partial_i A^i = 0 ,
\end{equation}
\noindent where the boundary indices $i,j$ are raised and lowered with $\eta_{ij}$. Eq. (\ref{Mr}) implies that we can use the residual gauge symmetry to set $\partial^i A_i = 0$ so that the remaining components become
\begin{equation}\label{Mi}
    r \partial^2_r A_i - (d-3)\partial_r A_i + r \Box_0 A_i = 0.
\end{equation}
Assuming harmonic dependence on the coordinates, for $d>2$ the general solution to (\ref{Mi}) takes the form
\begin{equation}\label{A k space}
    A_i(r,x) = e^{i k \cdot x} [ A^{(0)}_i(k) \psi_0 (k,r)  +  A^{(d)}_i(k) \psi_d (k,r) ]
\end{equation}
\noindent where
\begin{equation}\label{psi 0}
    \psi_0 = C_0(k,r) r^{d/2 - 1} Y_{d/2-1} (m_{bndy} r) = 1 + O(r^2) ,
\end{equation}
\begin{equation}\label{psi 1}
    \psi_d = C_d(k,r) r^{d/2 - 1} J_{d/2-1} (m_{bndy} r) =  r^{d-2} (1 + O(r^2) ),
\end{equation}
and $A^{(0)}$, $A^{(d)}$ are transverse form factors. The constants $C_0$, $C_d$ are chosen such that the asymptotics have the leading behavior displayed in (\ref{psi 0}), (\ref{psi 1}). For odd $d$, $\psi_d$ is an odd function of $r$ and $\psi_0$ is an even function of $r$ so that Neumann boundary conditions require $A^{(d)} = 0$. For even $d$, both $\psi_0$ and $\psi_d$ are even functions of $r$ (with some terms in $\psi_0$ containing $\ln r^2$) and in particular, both contain a term proportional to $r^{d-2}$. Here we define Neumann boundary conditions to be the choices of $A^{(0)}$ and $A^{(d)}$ for which the $O(r^{d-2})$ term cancels.

Let us first discuss the case of odd $d$, which is simpler as no logarithms arise. As for scalars, the most crucial step in the analysis is to determine allowed values of the boundary momentum $k^i$, assuming harmonic dependence $e^{ik\cdot x}$ on the boundary coordinates.  Solutions to (\ref{Mi}) with definite $k^i$ are Bessel functions, so for $k_ik^i$ not real normalizeability at the horizon requires a solution proportional to $K_{d/2-1}(p r)$ with $p^2 = k_i k^i$ and $\Re p > 0$.  But there are no such $p$ for which $K_{d/2-1}(p r)$ satisfies the Neumann condition.  On the other hand, there are modes with real timelike $k^i$ as well as lightlike modes (with profiles $1, r^{d-2}$).  The norms of these modes can be computed just as was done for scalars in section (\ref{Nnorm}).  For large $d$ this requires a large number of subtractions but, as noted for scalars,  one can use the argument from appendix D of \cite{Compere:2008us} to show that the role of these subtractions is always to precisely cancel the AdS boundary term in (\ref{rad int D PP SL}), leaving only a term at the horizon.  Just as in the scalar case, the lightlike modes are null states while the inner product of timelike modes is positive definite (for positive frequency). I.e., the results are just as for scalars with non-integer $\nu$.

We now turn to the case of even $d>2$.  Here, due to the appearance of logarithmic terms, the role of counter-terms is not simply to cancel the AdS boundary term in (\ref{rad int D PP SL}) and the renormalized SL inner product is more complicated.  However, because the relevant coefficients in the expansion of $K_{d/2-1}(p r)$ contain $\log p^2$ (whose range is the entire real line), this function satisfies our Neumann boundary condition for some real and positive value $p_0$.  Note that, since the corresponding momentum is spacelike, the transverse $(d-1)$-plane contains both timelike and spacelike vectors. Thus for any sign of the renormalized SL inner product, one of these polarizations must be a ghost.

Finally, consider the case $d=2$, where Neumann boudnary conditions are of particular interest. While for $d > 2$ the dual CFT contains a conserved current for Dirichlet boundary conditions, for $d=2$ the Dirichlet condition fixed the current to zero \cite{Marolf:2006nd}.  Only a Neumann-like condition can lead a conserved boundary current dual to a bulk Maxwell field (without Chern-Simons terms). For this reason, the Neumann theory was recently studied in \cite{Jensen:2010em}.  While the existence of tachyonic ghosts is clear from the fact that the magnetic dual of the Maxwell field in AdS${}_3$ is the massless ($\nu =1$) scalar studied in appendix (\ref{marginal}), it turns out to be interesting to see how the ghosts arise using the Maxwell description.

For $d = 2$, Maxwell fields have the asymptotic expansion
\begin{equation}\label{A asymp}
    A_i = A^{(1)}_i \log r^2 + A^{(0)}_i + \ldots  \,\,\,\,\, {\rm with} \,\,\,\, \partial^i A^{(1)}_i = 0
\end{equation}
Let us define Neumann boundary conditions by $A^{(0)} = 0$. So long as $A_r = O(1)$, the action
\begin{equation}\label{I d=2}
    I = I_0 + \int_{\partial M} A^{(1) i} A^{(1)}_i \log r^2 - \int_{\partial M} \sqrt{\gamma} f^i A_i  ,
\end{equation}
is finite and stationary \cite{Jensen:2010em}, where
\begin{equation}\label{fi}
    f^i := \rho_\mu F^{\mu i} = r^3 \eta^{ij} \partial_r A_j.
\end{equation}
In fact, a straightforward calculation yields $\delta I = 2 \int_{\partial M} A^{(0) i} \delta A^{(1)}_i$. The interesting point is that the counter-terms in (\ref{I d=2}) do not contain time derivatives.  Thus, the norm is given by the usual bulk expression
\begin{equation}\label{M Omega gral}
(A_1,A_2) = \frac{i}{2}  \int_\Sigma \sqrt{g_\Sigma} n_\mu ( F_1^{\mu \nu *}  A_{2 \nu} - F_2^{\mu \nu} A^*_{1 \nu}   )  = - \frac{i}{2} \int_\Sigma r^{3-d} [ A^{i}_2 \partial_t A_{1 i}^{*} - A^{i*}_1 \partial_t A_{2 i} ],
\end{equation}
where in the final step we chose $\Sigma$ to be a slice of constant $t$ and imposed exact radial gauge $A_r=0$ (so that we could use the conditions $A_r = \partial_i A^i = 0$).  Despite the fact that that no counter-terms were involved, there is nothing manifestly positive about the final expression in ({\ref{M Omega gral}) as the index $i$ is raised using the Lorentz-signature metric $\eta^{ij}$.  On the other hand, in a different gauge (in which the spatial part of the components $A_i$ along the boundary are purely transverse), the above norm {\it does} take a manifestly positive definite form (see e.g. equation (3.29) of \cite{Marolf:2006nd} in the gauge defined by setting their $A=0$). It was the norm in this second gauge that was used to analyze boundary conditions in \cite{Marolf:2006nd} and which led to the conclusion that the Neumann modes were not normalizeable.  The point here is that the `gauge transformation' required to transform between the above two gauges does not in general vanish at the boundary ($\partial \Sigma$).  As a result, it also contributes to the norm (and gives the term involving $A$ in equation (3.29) of \cite{Marolf:2006nd}).  We see that this gauge transformation\footnote{Here we abuse terminology in the usual way.  The fact that the transformation contributes to the norm means that it is not in fact pure gauge.} term plays the same role as the counter-terms introduced in treating scalar fields.

From this point the analysis proceeds as one expects (i.e., just as for the $\nu=1$ scalar in appendix \ref{marginal}).  The spectrum consists of time-like modes and a tachyon.  The norm of the time-like modes may be calculated as usual and is positive.  We now compute the norm of the tachyonic (T) solution $A_{i T} = e^{i k \cdot x} A^{(1)}_i \psi_T(z)$ where
\begin{equation}\label{psi tach}
    \psi_T(r) = - 2 K_0(p r) \approx \log r^2 + 2[\log(p/2) + \gamma] ,
\end{equation}
\noindent with $k_i k^i := p^2$ and $p := p_T = 2 e^{- \gamma}$ to satisfy the Neumann condition. The norm is
\begin{equation}\label{}
    (A_T, A_T) = \frac{1}{2}(w_1 + w_2) e^{it(w_1 - w_2)} (2 \pi) \delta(\vec{k}_1 -\vec{k}_2) (A^{(1) i}(k_1))^* A^{(1)}_i(k_2)) \langle \psi_{1 T}, \psi_{2 T} \rangle .
\end{equation}
The equation above is to be understood in the limit in which both momenta approach the value $p=p_T$ for the tachyon (in which case the time dependence cancels). The Sturm-Liouville product turns out to be positive:
\begin{equation}\label{}
    \langle \psi_{1 T}, \psi_{2 T} \rangle = \frac{2}{p^2_T} > 0.
\end{equation}
However, because the form factors $A^{(1)}_i$ are transverse and the momentum is space-like, the factor $(A^{(1) i}(k_1))^* A^{(1)}_i(k_2))$ is negative and the tachyon is a ghost.

Having found the spectrum, let us now calculate the 2-point function in position space. We consider only the ghost-free case of odd $d$.  We impose radial gauge and also guage-fix any lightlike modes to zero.   The norm of the timelike modes is
\begin{equation}\label{ip M tl}
   (A_1, A_2) =  \delta^{(d)}(k_1 - k_2) \frac{(2 \pi)^{d+1}}{2^d \Gamma(d/2-1)^2} m_{bndy}^{d-2} (A^{(0)}_i(k_1))^* A^{(0) i}(k_1)  \,\,\,\,\,\,\,\, {\rm for \ odd \  } d .
\end{equation}
Since the form factors are transverse and the momentum is time-like, it follows that $(A^{(0)}_i(k_1))^* A^{(0) i}(k_1) > 0$, so indeed (\ref{ip M tl}) is positive definite as expected. Now, in close analogy with section \ref{IR div}, we find
\begin{equation}\label{A 2pt x}
    \langle A_i (x_1,r_1) A_j (x_2,r_2) \rangle = \frac{2^d \Gamma(d/2-1)^2}{(2 \pi)^{d+1}}  \int_{\omega \ge |\vec k|} d \omega d^{d-1} \vec k  e^{ik \cdot(x_1-x_2)} \frac{\psi_0(r_1,k)\psi_0(r_2,k)}{(\omega^2 - |\vec k|^2)^{d/2-1}} P_{ij}   .
\end{equation}
\noindent where the projector $P_{ij} = \frac{1}{2} (\eta_{ij} + \frac{k_i k_j}{m^2_{bndy}})$ takes into account the fact that the gauge fields are transverse. Note that (\ref{A 2pt x}) diverges at $\omega = |\vec{k}|$ for $d\ge5$ (since $d$ is odd), as does the boundary propagator. However, the theory indeed exists for $d=3$ (where the lightlike modes are not normalizeable).


\begin{thebibliography}{10}

\bibitem{Maldacena:1997re}
J.~M. Maldacena, {\it The large {N} limit of superconformal field theories and
  supergravity},  {\em Adv. Theor. Math. Phys.} {\bf 2} (1998) 231--252,
  [\href{http://xxx.lanl.gov/abs/hep-th/9711200}{{\tt hep-th/9711200}}].

\bibitem{Witten:1998qj}
E.~Witten, {\it Anti-de {S}itter space and holography},  {\em Adv. Theor. Math.
  Phys.} {\bf 2} (1998) 253--291,
  [\href{http://xxx.lanl.gov/abs/hep-th/9802150}{{\tt hep-th/9802150}}].

\bibitem{Gubser:1998bc}
S.~S. Gubser, I.~R. Klebanov, and A.~M. Polyakov, {\it Gauge theory correlators
  from non-critical string theory},  {\em Phys. Lett.} {\bf B428} (1998)
  105--114, [\href{http://xxx.lanl.gov/abs/hep-th/9802109}{{\tt
  hep-th/9802109}}].

\bibitem{Aharony:1999ti}
O.~Aharony, S.~S. Gubser, J.~M. Maldacena, H.~Ooguri, and Y.~Oz, {\it Large {N}
  field theories, string theory and gravity},  {\em Phys. Rept.} {\bf 323}
  (2000) 183--386, [\href{http://xxx.lanl.gov/abs/hep-th/9905111}{{\tt
  hep-th/9905111}}].

\bibitem{Breitenlohner:1982bm}
P.~Breitenlohner and D.~Z. Freedman, {\it Positive energy in anti-de {S}itter
  backgrounds and gauged extended supergravity},  {\em Phys. Lett.} {\bf B115}
  (1982) 197.

\bibitem{Breitenlohner:1982jf}
P.~Breitenlohner and D.~Z. Freedman, {\it Stability in gauged extended
  supergravity},  {\em Ann. Phys.} {\bf 144} (1982) 249.

\bibitem{Balasubramanian:1998sn}
V.~Balasubramanian, P.~Kraus, and A.~E. Lawrence, {\it {Bulk vs. boundary
  dynamics in anti-de Sitter spacetime}},  {\em Phys. Rev.} {\bf D59} (1999)
  046003, [\href{http://xxx.lanl.gov/abs/hep-th/9805171}{{\tt
  hep-th/9805171}}].

\bibitem{Klebanov:1999tb}
I.~R. Klebanov and E.~Witten, {\it {AdS/CFT} correspondence and symmetry
  breaking},  {\em Nucl. Phys.} {\bf B556} (1999) 89--114,
  [\href{http://xxx.lanl.gov/abs/hep-th/9905104}{{\tt hep-th/9905104}}].

\bibitem{Witten:2001ua}
E.~Witten, {\it Multi-trace operators, boundary conditions, and {AdS/CFT}
  correspondence},  \href{http://xxx.lanl.gov/abs/hep-th/0112258}{{\tt
  hep-th/0112258}}.

\bibitem{Berkooz:2002ug}
M.~Berkooz, A.~Sever, and A.~Shomer, {\it Double-trace deformations, boundary
  conditions and spacetime singularities},  {\em JHEP} {\bf 05} (2002) 034,
  [\href{http://xxx.lanl.gov/abs/hep-th/0112264}{{\tt hep-th/0112264}}].

\bibitem{Gubser:2002zh}
S.~S. Gubser and I.~Mitra, {\it Double-trace operators and one-loop vacuum
  energy in {AdS/CFT}},  {\em Phys. Rev. D} {\bf 67} (2003) 064018,
  [\href{http://xxx.lanl.gov/abs/hep-th/0210093}{{\tt hep-th/0210093}}].

\bibitem{Gubser:2002vv}
S.~S. Gubser and I.~R. Klebanov, {\it A universal result on central charges in
  the presence of double-trace deformations},  {\em Nucl. Phys.} {\bf B656}
  (2003) 23--36, [\href{http://xxx.lanl.gov/abs/hep-th/0212138}{{\tt
  hep-th/0212138}}].

\bibitem{Hartman:2006dy}
T.~Hartman and L.~Rastelli, {\it Double-trace deformations, mixed boundary
  conditions and functional determinants in {AdS/CFT}},
  \href{http://xxx.lanl.gov/abs/hep-th/0602106}{{\tt hep-th/0602106}}.

\bibitem{Amsel:2008iz}
A.~J. Amsel and D.~Marolf, {\it {Supersymmetric Multi-trace Boundary Conditions
  in AdS}},  {\em Class.Quant.Grav.} {\bf 26} (2009) 025010,
  [\href{http://xxx.lanl.gov/abs/0808.2184}{{\tt arXiv:0808.2184}}].

\bibitem{Amsel:2009rr}
A.~J. Amsel and G.~Compere, {\it {Supergravity at the boundary of AdS
  supergravity}},  {\em Phys.Rev.} {\bf D79} (2009) 085006,
  [\href{http://xxx.lanl.gov/abs/0901.3609}{{\tt arXiv:0901.3609}}].

\bibitem{Witten:2003ya}
E.~Witten, {\it {$SL(2,Z)$} action on three-dimensional conformal field
  theories with {A}belian symmetry},
  \href{http://xxx.lanl.gov/abs/hep-th/0307041}{{\tt hep-th/0307041}}.

\bibitem{Marolf:2006nd}
D.~Marolf and S.~F. Ross, {\it {Boundary Conditions and New Dualities: Vector
  Fields in AdS/CFT}},  {\em JHEP} {\bf 0611} (2006) 085,
  [\href{http://xxx.lanl.gov/abs/hep-th/0606113}{{\tt hep-th/0606113}}].

\bibitem{Minwalla:1997ka}
S.~Minwalla, {\it Restrictions imposed by superconformal invariance on quantum
  field theories},  {\em Adv. Theor. Math. Phys.} {\bf 2} (1998) 781--846,
  [\href{http://xxx.lanl.gov/abs/hep-th/9712074}{{\tt hep-th/9712074}}].

\bibitem{Compere:2008us}
G.~Compere and D.~Marolf, {\it {Setting the boundary free in AdS/CFT}},  {\em
  Class.Quant.Grav.} {\bf 25} (2008) 195014,
  [\href{http://xxx.lanl.gov/abs/0805.1902}{{\tt arXiv:0805.1902}}].

\bibitem{Skenderis:2002wp}
K.~Skenderis, {\it {Lecture notes on holographic renormalization}},  {\em
  Class. Quant. Grav.} {\bf 19} (2002) 5849--5876,
  [\href{http://xxx.lanl.gov/abs/hep-th/0209067}{{\tt hep-th/0209067}}].

\bibitem{Grinstein:2008qk}
B.~Grinstein, K.~A. Intriligator, and I.~Z. Rothstein, {\it {Comments on
  Unparticles}},  {\em Phys.Lett.} {\bf B662} (2008) 367--374,
  [\href{http://xxx.lanl.gov/abs/0801.1140}{{\tt arXiv:0801.1140}}].

\bibitem{Grumiller:2008qz}
D.~Grumiller and N.~Johansson, {\it {Instability in cosmological topologically
  massive gravity at the chiral point}},  {\em JHEP} {\bf 07} (2008) 134,
  [\href{http://xxx.lanl.gov/abs/0805.2610}{{\tt arXiv:0805.2610}}].

\bibitem{Andrade:2009ae}
T.~Andrade and D.~Marolf, {\it {No chiral truncation of quantum log gravity?}},
   {\em JHEP} {\bf 03} (2010) 029,
  [\href{http://xxx.lanl.gov/abs/0909.0727}{{\tt arXiv:0909.0727}}].

\bibitem{Li:2008dq}
W.~Li, W.~Song, and A.~Strominger, {\it {Chiral Gravity in Three Dimensions}},
  {\em JHEP} {\bf 0804} (2008) 082,
  [\href{http://xxx.lanl.gov/abs/0801.4566}{{\tt arXiv:0801.4566}}].

\bibitem{Maloney:2009ck}
A.~Maloney, W.~Song, and A.~Strominger, {\it {Chiral Gravity, Log Gravity and
  Extremal CFT}},  {\em Phys.Rev.} {\bf D81} (2010) 064007,
  [\href{http://xxx.lanl.gov/abs/0903.4573}{{\tt arXiv:0903.4573}}].

\bibitem{Abramowitz1965}
M.~Abramowitz and I.~Stegun, {\em Handbook of Mathematical Functions}.
\newblock Dover (NY), 1965.

\bibitem{Faulkner:2010gj}
T.~Faulkner, G.~T. Horowitz, and M.~M. Roberts, {\it {Holographic quantum
  criticality from multi-trace deformations}},  {\em JHEP} {\bf 1104} (2011)
  051, [\href{http://xxx.lanl.gov/abs/1008.1581}{{\tt arXiv:1008.1581}}].

\bibitem{Hawking:2000bb}
S.~Hawking, T.~Hertog, and H.~Reall, {\it {Trace anomaly driven inflation}},
  {\em Phys.Rev.} {\bf D63} (2001) 083504,
  [\href{http://xxx.lanl.gov/abs/hep-th/0010232}{{\tt hep-th/0010232}}].

\bibitem{Bakas:2008gz}
I.~Bakas, {\it {Energy-momentum/Cotton tensor duality for AdS(4) black holes}},
   {\em JHEP} {\bf 0901} (2009) 003,
  [\href{http://xxx.lanl.gov/abs/0809.4852}{{\tt arXiv:0809.4852}}].

\bibitem{deHaro:2008gp}
S.~de~Haro, {\it {Dual Gravitons in AdS(4) / CFT(3) and the Holographic Cotton
  Tensor}},  {\em JHEP} {\bf 0901} (2009) 042,
  [\href{http://xxx.lanl.gov/abs/0808.2054}{{\tt arXiv:0808.2054}}].

\bibitem{Mansi:2008bs}
D.~S. Mansi, A.~Petkou, and G.~Tagliabue, {\it {Gravity in the 3+1-Split
  Formalism II: Self-Duality and the Emergence of the Gravitational
  Chern-Simons in the Boundary}},  {\em Class.Quant.Grav.} {\bf 26} (2009)
  045009, [\href{http://xxx.lanl.gov/abs/0808.1213}{{\tt arXiv:0808.1213}}].

\bibitem{Leigh:2003gk}
R.~G. Leigh and A.~C. Petkou, {\it Holography of the {$N = 1$} higher-spin
  theory on {AdS$_4$}},  {\em JHEP} {\bf 06} (2003) 011,
  [\href{http://xxx.lanl.gov/abs/hep-th/0304217}{{\tt hep-th/0304217}}].

\bibitem{Leigh:2003ez}
R.~G. Leigh and A.~C. Petkou, {\it {$SL(2,Z)$} action on three-dimensional
  {CFT}s and holography},  {\em JHEP} {\bf 12} (2003) 020,
  [\href{http://xxx.lanl.gov/abs/hep-th/0309177}{{\tt hep-th/0309177}}].

\bibitem{Bakas:2008zg}
I.~Bakas, {\it {Duality in linearized gravity and holography}},  {\em
  Class.Quant.Grav.} {\bf 26} (2009) 065013,
  [\href{http://xxx.lanl.gov/abs/0812.0152}{{\tt arXiv:0812.0152}}].

\bibitem{ElShowk:2011gz}
S.~El-Showk, Y.~Nakayama, and S.~Rychkov, {\it {What Maxwell Theory in D$\neq$4
  teaches us about scale and conformal invariance}},  {\em Nucl. Phys.} {\bf
  B848} (2011) 578--593, [\href{http://xxx.lanl.gov/abs/1101.5385}{{\tt
  arXiv:1101.5385}}].

\bibitem{Ishibashi:2003jd}
A.~Ishibashi and R.~M. Wald, {\it Dynamics in non-globally-hyperbolic static
  spacetimes. {II}: General analysis of prescriptions for dynamics},  {\em
  Class. Quant. Grav.} {\bf 20} (2003) 3815--3826,
  [\href{http://xxx.lanl.gov/abs/gr-qc/0305012}{{\tt gr-qc/0305012}}].

\bibitem{Weinberg:1996kr}
S.~Weinberg, {\it {The quantum theory of fields. Vol. 2: Modern applications}},
  .

\bibitem{Heemskerk:2010hk}
I.~Heemskerk and J.~Polchinski, {\it {Holographic and Wilsonian Renormalization
  Groups}},  \href{http://xxx.lanl.gov/abs/1010.1264}{{\tt arXiv:1010.1264}}.

\bibitem{Faulkner:2010jy}
T.~Faulkner, H.~Liu, and M.~Rangamani, {\it {Integrating out geometry:
  Holographic Wilsonian RG and the membrane paradigm}},
  \href{http://xxx.lanl.gov/abs/1010.4036}{{\tt arXiv:1010.4036}}.

\bibitem{Jensen:2010em}
K.~Jensen, {\it {Chiral anomalies and AdS/CMT in two dimensions}},  {\em JHEP}
  {\bf 01} (2011) 109, [\href{http://xxx.lanl.gov/abs/1012.4831}{{\tt
  arXiv:1012.4831}}].

\end{thebibliography}

\end{document}